\documentstyle[prd,aps,epsfig,floats]{revtex}
\begin{document}
\draft
\wideabs{
\title{A Regularization of Quantum Gravity}
\author{Wolfgang Beirl and Bernd A. Berg} 
\address{ (E-mails: wb1@gate.net, berg@hep.fsu.edu)\\ 
Department of Physics, The Florida State University,
  Tallahassee, FL 32306} 
\date{March 8, 2001}
\maketitle
\begin{abstract}
We re-examine results of the Liouville theory and provide arguments
that a {\it negative} bare cosmological constant is essential to define
two-dimensional quantum gravity. From this we are naturally led to a
regularization of quantum gravity within the Regge approach such that
it is described by small fluctuations around equilateral triangles,
whose average link length approaches zero in the continuum limit. We
investigate a model based on this idea numerically and present evidence
for the desired long-range correlations. Interestingly, the approach 
might generalize to higher dimensions. The picture of an inflated 
balloon, which is often used to demonstrate the properties of an 
expanding classical universe, seems to be valuable to understand 
quantum gravity as well.
\end{abstract}
}

\narrowtext

\section{Introduction} \label{sec_intro}

In recent years, substantial progress has been achieved in
understanding two-dimensional quantum gravity based on the Liouville
theory~\cite{Polyakov}, see~\cite{Hatfield} for a review.
Unfortunately it was not possible to extend these results to higher
dimensions. In the light of these difficulties we re-examine the results
of Polyakov, starting with the path integral of two-dimensional
gravity for a fixed topology and with\footnote{Following established
notation~\cite{Hatfield}, in the present context $D$ denotes the
number of matter fields or dimension of the embedding space.} D additional
matter fields $X^i$:
\begin{eqnarray} \nonumber
 Z_D(\mu_0) = \int {\cal D} g\, {\cal D} X^i\, \times 
~~~~~~~~~~~~~~~~~~~~~~~~~~~~~~~\\
\exp{ \left( -\mu_0 \int d^2x \sqrt{g}
- \int d^2x \sqrt{g} g^{ab} \partial_a X^i \partial_b X^i \right) }\ .
\label{equ1}
\end{eqnarray}
Following the procedure of Polyakov, we 'fix a gauge' using coordinates
$x$ so that $g_{ab} = e^\phi h_{ab}$, where $h$ is an external 
background metric. Using an appropriate regularization, this leads to 
the well-known path integral
\begin{eqnarray} \label{Z_Liouville}
 Z_D(\mu_s) = \int {\cal D} \phi (x)\, \times
~~~~~~~~~~~~~~~~~~~~~~~~~~~~~~~~~\\ \nonumber
 \exp{ \left[ - \frac{26-D}{48\pi}
 \int d^2x\, \left( {1\over 2}\, (\partial \phi)^2 +
 \mu_s e^\phi + R \phi \right) \right] }
\end{eqnarray}
for the Liouville field $\phi(x)$, with $R$ being the curvature of the
metric $h$, which is chosen so that $R$ is a constant depending on the
topology~\cite{Polyakov}. The conformal anomaly is present for
$D \neq 26$ where a scalar field appears in the quantum theory
although there are no physical degrees of freedom in the classical
theory. (There are no field equations for gravity in two dimensions.)
The parameter $\mu_s$ is not equal to the bare cosmological constant
$\mu_0$ and the evaluation of the necessary counter-terms
reveals~\cite{epsilon}
\begin{equation} \label{mu0}
\mu_0=\frac{D-2}{8\pi\epsilon}+\mu_s\,\left(\frac{26-D}{48\pi}\right)
\label{equ3} \end{equation}
where $\epsilon$ is the cut-off parameter of the
regularization procedure. Usually
the path integral~(\ref{Z_Liouville}) is the starting point
for further examination. We emphasize only that for
$R<0$ the interaction term $R\phi + \mu_s e^\phi$ has a true
minimum at $\phi_0=\ln(-R/\mu_s)$ and using $\phi(x)=\phi_0+\eta(x)$
one can approximate the vacuum functional by
\begin{eqnarray} \label{Ieta}
 Z_D(\mu_s) = \int {\cal D} \eta\,  \times
~~~~~~~~~~~~~~~~~~~~~~~~~~~~~~~~~~~~~~\\  \nonumber
 \exp{ \left[ - \frac{26-D}{48\pi}
 \int d^2x\, \left( {1\over 2}\, (\partial \eta)^2 +
 ( c_0 \eta^2 + c_1 \eta^3 + ... ) \right) \right] }
\end{eqnarray}
with $c_0 = - \frac{1}{2}R$ etc.~\cite{Triest}.

We highlight the fact that for $D < 2$  
relation~(\ref{mu0}) indicates that the bare cosmological constant
$\mu_0$ is negative and infinite in the limit $\epsilon \to 0$.
While the path integral~(\ref{equ1}) seems to be highly divergent for
$\mu_0 \to -\infty$, the Polyakov procedure transforms it so that 
the scalar Liouville field results in two dimensions.
In this paper we want to understand how this is possible, using
the framework of the Regge calculus~\cite{Regge}. 
Subsequently, we touch briefly on the prospects of generalizing our 
regularization procedure to higher dimensions.
For our investigations of quantum gravity in the Regge regularization 
we use a simple measure, thus defining a model.  Although it is not 
clear whether this model can be fully equivalent to the Liouville 
theory in two dimensions, it shares some of its remarkable
properties and exhibits a phase transition which appears to define a
continuum theory. However, it might be that two dimensional gravity
is a poor testing ground for the general case~\cite{BB95}.
In higher dimensions the Einstein Hilbert action
should dominate the physics of quantum gravity and since, without 
proper regularization, it is unbounded in the Euclidean sector,
its behavior needs to be determined before one can discuss the
subtleties of the correct measure.

In the next section we define the model and obtain analytical results
including properties of its phase transition. Numerical simulations 
are subsequently performed in section~\ref{simulations}. They yield
long-range correlations as needed for the continuum limit. Conclusions
follow in section~\ref{conclusions}.

\section{The model}

For $D=0$ we can write the path integral (\ref{equ1}) as
\begin{equation} \label{Z_gravity}
Z(\mu_0) = \int Dg\, e^{ - \mu_0 A[g] }
\end{equation}
where $A[g]$ is the total area of the geometry. The integral
(\ref{Z_gravity}) is defined only after proper regularization
as the following example~\cite{Ambjorn} illustrates:
Consider a smooth geometry of finite total area. Then, assume
that a "spike" of length $L$ and circumference $u$ grows
out of this area. The spike can have arbitrarily large length $L$,
but, if $u$ is sufficiently small, it would not be suppressed by the
action of the path integral due to its infinitesimal area $\sim uL$.
Indeed, the path integral would be dominated by ill-defined
geometries, unless we introduce a regularization procedure, which
provides for a cutoff to prohibit $L$ from becoming arbitrarily large.
A proper regularization procedure introduces counter-terms, which 
prevent degenerate geometries and lead to a well-defined path
integral~\cite{Polyakov}, as we will discuss in the following
for a model of pure gravity.

We consider a triangulation of fixed topology (e.g a 2-torus),
with $N_2$ triangles and $N_1$ links. We assume that the triangulation 
is sufficiently regular and, following the Regge approach, consider the
link lengths $x$ as the variables of this model. We use a cutoff $a$ on
the link lengths and the simple local measure $\prod_l d x_l$
to obtain the path integral
\begin{equation} \label{Z_local}
Z(\mu_0,a) = \int_0^a \prod_l dx_l\, e^{ - \mu_0 A[x] }
\label{eq6}
\end{equation}
where the total area $A[x]$ of the geometry is the sum of all
triangle areas,
$$ A = \sum_t A_t\ ,$$
and the integration range restricts the link lengths to $0<x<a$. The
above integral is obviously well-defined for all $\mu_0$ as long as
$a$ is finite.

It is not clear whether the partition function~(\ref{Z_local})
can be in the same universality class as~(\ref{Z_gravity}), i.e. 
that diffeomorphism invariance will be properly implemented by our
simple measure in a continuum limit.  It is known, however, that the
Regge approach approximates Einstein gravity in the classical 
limit~\cite{classical_limit}.
We reach the continuum limit formally by increasing the number of
links $N_1$ and decreasing their lengths, $a \to 0$.

To proceed further, we re-write the path integral (\ref{Z_local}) as
\begin{equation} \label{Z_local2}
Z(\mu_0,a,n) = \int_0^\infty \prod_l dx_l\,
 \exp{ \left( - \mu_0 A[x] - \sum_l (x_l/a)^n \right) }
\end{equation}
i.e. replacing the cutoff with a counter-term, which is equivalent
in the limit $n\to \infty$.

Let us qualitatively examine how the Regge lattice might behave
for different values of $\mu_0$. A positive cosmological constant, 
$\mu_0 > 0$, tends to suppress large areas, while the counter-term
restricts the link lengths. In the limit $\mu_0 \to +\infty$ the
expectation value of the total area will tend to $0$, but it is not
clear whether the link lengths will approach $0$ as well. It is
possible that their expectation value remains finite and this would
result in a 'crumpled' lattice with collapsed triangles. The numerical
simulations presented in the following section suggest
 that this indeed
happens and it is not clear if a reasonable continuum limit can be
found for $\mu_0 > 0$. Previous numerical
investigations~\cite{HamberJanke} have been mostly confined to this
range of a positive cosmological constant.

On the other hand, if $\mu_0 < 0$, the cosmological term tends to
inflate the area of the lattice, while the counter-term limits the
link lengths. Therefore, equilateral triangles are preferred and
for $\mu_0 \to -\infty$ we have to expect a lattice which consists
of (almost) equilateral triangles, with small fluctuations around
the maximum link length $a$; other configurations cannot significantly
contribute to the path integral.
The situation is thus similar to an inflated balloon, where the
pressure of the air is the analogue to the cosmological term,
while the role of the rubber molecules is played by the Regge
links. If we inflate the balloon enough, the rubber becomes
locally flat and elastic. While it is immediately clear that
the Regge lattice is locally flat for regular triangulations in two 
dimensions~\cite{flat}, if $\mu_0$ is sufficiently
negative, we must examine in the following whether the lattice fluctuations
are able to reproduce a scalar field as in the Liouville theory.
To this purpose we consider first the path integral for
$\mu_0 < 0$ and $n > 2$. The special case of $n=2$ is
then discussed separately.

\subsection{$n>2$} \label{ngt2}
 
If $n$ is large but finite and $\mu_0$ is sufficiently negative, 
only lattice configurations with link lengths $x \sim a$  
contribute significantly to the path integral. Configurations with
$x >> a$ are suppressed by the term $- (x/a)^n$, while configurations
with $x << a$ cannot contribute significantly due to the area term
$-\mu_0\,A>0$. We thus replace $x_l$ with $a(1 + \xi_l)$ where the
variable $\xi_l$ determines the fluctuation of link $l$ around the
length $a$. The path integral becomes
\begin{eqnarray} \nonumber
Z(\mu_0,a,n) = a^{N_1} \int \prod_l d\xi_l\, \times
~~~~~~~~~~~~~~~~~~\\
 \exp{ \left( - \mu_0 a^2
 \sum_t A_t[1+\xi] - \sum_l (1+\xi_l)^n \right) }
\label{Z_1} \end{eqnarray}
and only configurations $\xi \sim 0$ contribute significantly.
Notice that the cut-off $a$ appears as a simple, unphysical
multiplicator of the bare cosmological constant, which allows
us to remove it from the path integral by rescaling $\mu_0$.
 
The area of a single triangle with the links $a, b, c$ is of the form
\begin{eqnarray} \label{Area_1}
A_t[1+\xi] &=& A_0 + A_1(\xi_a + \xi_b + \xi_c) -
A_2(\xi_a^2 + \xi_b^2 + \xi_a^2) \nonumber \\
&-& B_2 ((\xi_a - \xi_b)^2 + (\xi_a - \xi_c)^2  \nonumber \\
&+& (\xi_b - \xi_c)^2) + O(\xi^3) 
\end{eqnarray}
with positive constants
\begin{eqnarray} \label{t_constants}
 A_0={\sqrt{3}\over 4},\  A_1={1\over 2\sqrt{3}},\
 A_2={1\over 12\sqrt{3}}\ {\rm and}\ B_2={1\over 6\sqrt{3}}.
\end{eqnarray}
The cosmological constant term, i.e. the total area of the geometry,
can therefore be expanded as
\begin{eqnarray} \nonumber
\sum_tA_t[1+\xi] &=& A_0N_2 + 2A_1\sum_l\xi_l - 2A_2\sum_l\xi_l^2\\
&-& B_2\sum_{[mn]} (\xi_m - \xi_n)^2 + O(\xi^3)
\label{Appr_2} \end{eqnarray}
where the sum over $[nm]$ indicates a sum over neighboring links. 
Correspondingly we expand the regulating link term as
\begin{equation} \label{Appr_3}
\sum_l(1+\xi_l)^n = N_1 + n\sum_l\xi_l +
 \frac{n(n-1)}{2}\sum_l\xi_l^2 + O(\xi^3)
\end{equation}
and insert both expansions in the path integral. We assume equilibrium
of the link lengths around $a$ and for consistency we have to set 
$\mu_0 a^2 2A_1 = -n$, so that the first order term in the action 
vanishes.  This determines the bare cosmological constant as 
\begin{equation} \label{mu_0}
\mu_0=- {n\over 2 a^2 A_1}
\end{equation}
so that it is negative and tends to $-\infty$ for 
$a \to 0$. (Notice the similarity with relation~(\ref{mu0}).)

The path integral becomes    
$$ Z(n) = const \int \prod_l d\xi_l\, \exp \left[\, - \left( 
 \frac{n(n-1)}{2} + n \frac{A_2}{A_1} \right)\, \sum_l \xi_l^2
 \right. $$
\begin{equation}
  - \left. n \frac{B_2}{A1} \sum_{mn} 
  ( \xi_m - \xi_n )^2 + O(\xi^3)\, \right] 
\label{Z_2} \end{equation}
where we now understand $n$ as a coupling parameter which is not
necessarily an integer. While the first term guarantees that
$|\xi|\ll 1$ for large enough $n$, the second term can obviously be
interpreted as the discretization of the kinetic term of a scalar
field $\xi$, similarly to the field $\eta$ of the path
integral~(\ref{Ieta}).

To proceed further, we discretize the variables $\xi_l$
using $\epsilon \sigma_l$ instead, with $\sigma_l$ being an integer
$\pm 1, \pm 3, \pm 5, ...$, so that the difference between two
neigbouring values of $\xi$ equals $2\epsilon$. The path integral
is then replaced by a sum over different configurations $\sigma_l$
\begin{eqnarray}  \nonumber
Z(\beta,\epsilon) = \sum_{[\sigma]} \exp \left(\, -
\beta^2 \epsilon^{-2} \sum_l \sigma_l^2 \right. \\
\left.  - \beta \frac{B_2}{A1} \sum_{mn} ( \sigma_m - \sigma_n )^2 
+ O(\epsilon^3) \right)
\label{Z_sum} \end{eqnarray}
where we use $\beta = n \epsilon^2$ and assume that $n$ is large
enough that the coupling of the first term in (\ref{Z_2}) is
essentially $n^2/2$.

The limit $\epsilon \to 0$ leads back to the path integral (\ref{Z_2})
and, if we adjust $n$ such that $\beta$ remains finite, the
$\beta^2\epsilon^{-2}$ term of (\ref{Z_sum}) will suppress all higher
values of $\sigma_l$. It is then sufficient to perform a summation of
all configurations $\sigma_l = +1, -1$ and the system becomes equivalent
to an Ising model. Therefore, in this limit the model exhibits a second
order phase transition at a certain critical coupling $\beta_c$, which
leads us to the equivalence with an interacting scalar field with
$S$-matrix $-1$~\cite{Japanese}. Although this field is in a different 
universality class than the Liouville theory, the associated long-range 
correlations demonstrate that our 'quantum balloon' is indeed elastic 
as we had hoped for. In previous investigations it has been proposed to
approximate quantum gravity by Ising models and numerical simulations
of such models have been performed~\cite{Vienna}. Our calculation
demonstrates that these Ising models follow naturally in a certain
limit ($\mu_0 \to -\infty, n\to \infty$) of the Regge approach.

The occurance of different universality classes is certainly related to
the sensitive dependence of two dimensional quantum gravity on the
definition of the measure. 
The above calculation is supposed to depend on the details of the
limit ($\mu_0 \to -\infty, n\to \infty$) and, furthermore, we also do 
not know the universality classes for general $n$.  In principle, this 
could be determined by calculating the critical exponents numerically. 
So far, we did not investigate this issue in detail. It is expected to 
be of minor importance in higher dimensions due to the presence and 
dominance of the Einstein-Hilbert term there.

\subsection{$n=2$}  \label{neq2}
  
Having discussed the case $n>2$ and a particular limit $n\to\infty$, we
examine the special case $n=2$ in the following. The reason for our
interest in this case is that the critical point is found for a finite
value of $\mu_0$, which allows for easier numerical investigations.

We use the dimensionless variables
\begin{equation} \label{q_l}
q_l = (x_l/a)^2
\end{equation}
with the measure $\prod_l dq_l$ instead of $\prod_l d x_l$
for reasons of analytical simplicity. Using the rescaled coupling
parameter
\begin{equation} \label{mu}
 \mu = \mu_0 a^{-2}
\end{equation}
the path integral (\ref{Z_local2}) becomes
\begin{equation} \label{Zq1}
Z(\mu)=\int_0^{\infty} \prod_l dq_l\,
\exp \left(-\mu A[q]-\sum_l q_l\right)\ .
\end{equation}
This path integral is well defined as long as $\mu > \mu_c$ and the 
value of the critical coupling $\mu_c$ is found from the
configuration with all link lengths equal, $q_l = q$. The action
reduces to $( -\mu (\sqrt{3}/2) N_2 - N_1 )q$ in this case and the 
exponent remains finite only if $\mu>\mu_c=-(2/\sqrt{3})(N_1/N_2)$.
In two dimensions we have $N_1 = 3N_2/2$ and
\begin{equation} \label{mu_c}
 \mu_c = - \sqrt{3}
\end{equation}
is independent of the lattice size. 
Although the lattice is expected to fluctuate heavily at the critical point,
we have nevertheless determined $\mu_c$ by assuming that
configurations around equilateral triangles dominate the path integral.
Unfortunately, we cannot prove this assumption, but it is confirmed by
our numerical simulations (see the next section) and it is
self-consistent as we show in the following.

We substitute the $N_1$ variables $[q_1, ..., q_{N_1}]$ by the $N_1$ 
variables $[q, \xi_1, ..., \xi_{N_1 - 1}]$ defined by
\begin{equation} \label{substitute_1}
 q_1 = q(1 + \xi_1),\ q_2 = q(1 + \xi_2),\ \dots
\end{equation}
\begin{equation} \label{substitute_2}
{\rm and}~~~ q_{N_1}=q\, (1-\sum_{l=1}^{N_1-1}\xi_l)\, ,
\end{equation}
so that $\sum_l q_l=N_1 q$.
The total area $A[q]$ can then be written as 
\begin{equation} \label{tau}
 A[q]\ =\ (\sqrt{3}/2)\, N_2\, q\, \tau[\xi]
\end{equation}
where the function $\tau$ depends on the variables $\xi_l$ only and varies
between $0$ and $1$. It equals $1$ if all triangles are equilateral
(maximum area) and it is $0$ if all triangles collapse (minimum area). 
The path integral becomes
$$ Z(\mu) = \int_0^{\infty} dq\,q^{N_1-1}\, \times $$
\begin{equation} \label{Zqx}
\int \prod_l d\xi_l\, \exp{ \left(
- \frac{3}{2}\,\mu\,N_2\,q\,\tau[\xi] - N_1\,q \right) } 
\end{equation}
because the functional determinant is a constant, when it is evaluated 
under the assumption of small fluctuations around equilateral triangles, 
which allows us to ignore the triangle constraints.
Integrating out $q$, one arrives at 
\begin{equation} \label{Zt}
Z(\mu) = \frac{\Gamma(N_1)}{N_1^{N_1}} \int \prod_l d\xi_l\, \frac{1}{
(\,1+\frac{\sqrt{3}}{2}\,\mu\,\frac{N_2}{N_1}\, \tau[\xi]\,)^{N_1}}\ .  
\end{equation}
We are interested to determine the expectation value of
\begin{equation} \label{tau_limit}
<\tau>~~ {\rm for}~~
\mu \to \mu_c = - \frac{2}{\sqrt{3}}\frac{N_1}{N_2}
\end{equation}
and obviously the behavior of the function  $\tau[\xi]$ near $\tau = 1$ 
is essential in this case. We use the fact that $\tau$ has a maximum at
$\xi = 0$ and expand it as
$$ \tau[\xi] = 1 - B N_1^{-1} \sum_l \xi_l^2 -
 \sum_{[ij]} B_{ij} \xi_i \xi_j +O( \xi^3 ) $$
with a constant $B > 0$ and a constant matrix $B_{ij}$.
In a second step we introduce spherical coordinates
$\xi_i = r \sin \phi_1 \sin \phi_2 ...$ and thus arrive at 
\begin{equation} \label{Zr}
Z(\mu) = const \int \prod d\phi\, dr\, r^{N_1 - 2}\, \times
\end{equation}
$$
\frac{ J(\phi ) }{(\,1+\frac{\sqrt{3}}{2}\,\mu\,\frac{N_2}{N_1}\,
 [1-B\,r^2-r^2 \sum_{[ij]} B_{ij}\,f_{ij}(\phi)]\, )^{N_1}} 
$$
if we neglect terms of order $O(\xi^3)$. The function $J( \phi )$
denotes the Jacobian for the spherical coordinates and $f_{ij}$
denotes  a matrix of functions which depend on the angles $\phi$ only
such that $\xi_i \xi_j = r^2 f_{ij}( \phi )$.
At $\mu = \mu_c$ the path integral becomes 
$$ Z(\mu) = const \int \prod d\phi\, \frac{J(\phi )}{
(B+\sum_{[ij]} B_{ij}\,f_{ij}(\phi)\,)^{N_1}}\, \times $$
\begin{equation} \label{Zra}
 \int dr\, \frac{ r^{N_1 - 2} }{ r^{2 N_1} } 
\end{equation}
which we can write as $ const \int dr\, r^{ - (N_1 + 2) }$
in order to calculate the expectation value $<r>$. 
Obviously the path integral has a strong singularity at $r = 0$ 
and the expectation value of $r$ is zero at $\mu_c$. This, of course,
implies that $<\tau> = 1$.

In order to determine $<r>$ for $\mu = \mu_c + \delta$, the first
momentum $\int dr f(r)\, r / \int dr f(r)$ needs to be calculated
for the function
$$ f(r) = r^{N_1 - 1} / ( \delta + B' r^2 )^{N_1}\, ,$$ 
where $B'$ equals $B + \sum_{[ij]} f_{ij}[\phi]$.
The function $f$ has a maximum at $r_{\max} =
\sqrt{(N_1-1) \delta / (N_1B'+B')}$
and falls off sharply away from it. We conclude that
$<r> \sim const\, \delta^{\frac{1}{2}}$ where the constant depends on
the integration of $B'$ over the spherical angles $\phi$. Therefore,
\begin{equation} \label{tau_mu}
<\tau>\ \sim\ 1 - const\, (\mu - \mu_c)^{\frac{1}{2}}  
\end{equation}
for lattice configurations fluctuating around equilateral triangles,
providing for the self-consistency of our initial assumption.

An eventual, non-trivial continuum limit $\mu\to\mu_c$ is quite subtle.
The numerical investigations of section~\ref{simulations}
suggest the following picture:  
At $\mu_c$ the path integral is dominated by small fluctuations around 
equilateral triangles. The link lengths show correlations which for 
$\mu\to\mu_c$ are consistent with a diverging correlation length (mass 
to zero).  Considering the ratio $r=\min(x_l)/\max(x_l)$, one finds
that the expectation value $<r>$ shows a strong finite size effect,
in contrast to $<\tau>$ which is essentially independent of $N_1$.
The continuum limit is then found for $(N_1 \to \infty, \mu \to \mu_c)$,
so that $<\tau> \to 1$ and $<r> \to 0$, which indicates that the triangles
are (almost) equilateral yet over large distances the geometry of a
typical lattice is non-trivial.  

\subsection{Higher dimensions} \label{higher_d}

Let us briefly comment on the situation in higher dimensions. We focus 
on the physically interesting case of four dimensions and, because of
the importance of numerical calculations, on $n=2$. We consider the
path integral
\begin{equation} \label{Zq4d}
Z(\mu) = \int Dq\, \exp{ \left( - \mu V[q] - \sum_l q_l^2 \right) } 
\end{equation}
with total 4-volume $V[q]$ and a regulating counter-term. We use again
a simple, local measure for simplicity and we ignore the Einstein-Regge
action as well as other possible terms in the action.
We introduce variables $p$ and $\xi$ so that 
$p N_1 = \sum_l q_l^2$ and $q_i = p^{\frac{1}{2}} ( 1 + \xi_i )$ 
and we use the function $\nu[\xi]$ as we used $\tau$ for two dimensions, 
so that $V[q] = \sum_s V_s[q] = p \nu[\xi] V_0 N_4$
where the constant $V_0$ is chosen so that the maximum of $\nu$ equals $1$.
The path integral becomes 
\begin{eqnarray} \nonumber
Z(\mu) = \int_0^{\infty} dp\, p^{(N_1-2)/2}\, \times 
~~~~~~~~~~~~~~~~\\ \label{ZFNU4d}
 \int \prod_l d\xi_l\,  
 \exp \left( - \mu N_4 V_0 p \nu[\xi] - N_1 p \right)\ . 
\end{eqnarray}
Proceeding along the same lines as for the two-dimensional case, one
concludes that $< \nu > \to 1$ as $\mu$ approaches the critical value
$\mu_c$. At the critical point lattice configurations with 
equilateral 4-simplices contribute significantly only, indicating
long-range correlations and a non-trivial phase transition as in two
dimensions.

Unfortunately, a regular triangulation with equilateral simplices in 
four dimensions does not provide for a locally flat geometry. Also, 
the ratio $N_4/N_1$ is not constant and some properties of our 
regularization depend on the details of the triangulation. Clearly, a
mechanism is needed which would favor locally flat geometries, 
independent of the triangulation and we hope that the Einstein-Hilbert 
action will achieve this.
Indeed we can show that our regularization is sensitive to the 
Einstein-Hilbert action
near the critical point, by adding 
\begin{equation} \label{EH}
 S_{EH}(q) = \sum_t A_t \delta_t
\end{equation}
to the action in equation~(\ref{Zq4d}).  Using the gravitational 
coupling $\beta$ we write 
\begin{equation} \label{beta_EH}
\beta S_{EH}(q) = - \beta p^{1/2} s(\xi)
\end{equation}
where $s(\xi)$ is bounded because the deficit angles 
$-n\pi < \delta_t < 2\pi $ are bounded ($n$ depends on the details of 
the triangulation) for a given triangulation.

One can expand the action for small $\beta$ as 
$$\exp(-\beta p^{1/2} s(\xi)) = 1-\beta p^{1/2} s(\xi)+O(\beta^2)$$
and evaluate the path integral equation~(\ref{ZFNU4d}) with this 
correction. Performing the integration of the variable $p$ one finds 
that the integrand in the remaining path integral is modified by a 
factor
$$[\,1-\beta\,C(N_1)\,s(\xi_l)\,]\,/\,[\,1-(\mu/\mu_c)\,\nu(\xi_l)\,]$$
where $C(N_1)$ is a constant, which depends slightly on $N_1$. Near 
the critical point we can set $\xi = 0$ to approximate $\beta_c$
as the gravitational coupling which leads to
$$[\,1-\beta_c\,C(N_1)\,s(0)\,]\,/\,[\,1-(\mu/\mu_c)\,\nu(0)\,]=0\,.$$
Of course, this does not imply that $Z = 0$ at $\beta_c$, rather
we have to conclude that our approximation for small $\beta$ breaks down.    
It is now important to see that $\beta_c \to 0$ for $\mu \to \mu_c$,
which means that our model becomes very sensitive to the Einstein-Hilbert
term near the critical point. Numerical simulations could clarify the
relevance of this effect and in general the phase structure for the
two coupling parameters $\mu$ and $\beta$. This is left as a project
for future investigations.

\section{Numerical results} \label{simulations}

To support the assumptions as well as the results of our
analytical calculations in the previous section, we perform numerical 
simulations of our two dimensional model. We use the path integral
(\ref{Z_local2}) for $n=2$ on a 2-torus. Our consideration which leads
to $\mu_c=-\sqrt{3}$ of equation~(\ref{mu_c}) is not affected by the
change of the measure from the path integral (\ref{Zq1}) to
(\ref{Z_local2}).
We have generated data on $8\times 8$,
$16\times 16$ and $32\times 32$ lattices at $\mu = 5.0,\, 2.0,\,
1.0\, 0.0\, -1.0\, -1.5\, -1.7$ and $-1.72$. Each data point relies
on at least 1M sweeps and measurements are taken every 10 sweeps to reduce auto-correlation of the data.

\begin{figure}[-t] \begin{center}
\epsfig{figure=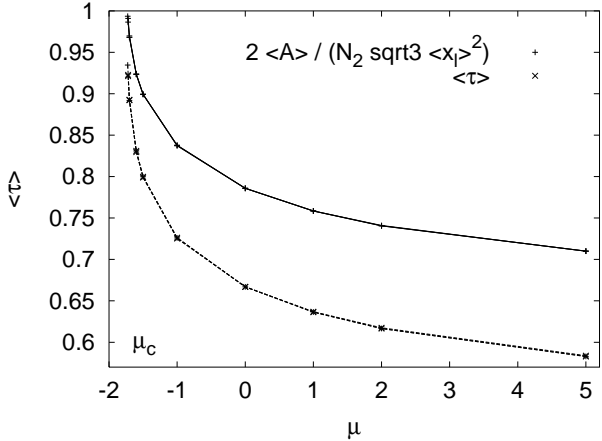,width=\columnwidth} \vspace*{-1mm}
\caption{Together with $2 \langle A\rangle / (N_2\sqrt{3} \langle 
x_l\rangle^2)$ the ratio $\langle \tau \rangle = 2 \langle A/x_l^2 
\rangle / (N_2\sqrt{3})$, defined by equation~(\ref{tau}), is depicted 
as function of the cosmological constant $\mu$. Within the accuracy 
of the plots data from $8\times 8$, $16\times 16$ and $32\times 32$ 
lattices fall on top of one another. \label{fig_tau} }
\end{center} \end{figure}

In figure~\ref{fig_tau} the expectation value
$$ \langle \tau \rangle = 2 \langle A / x_l^2 \rangle / (N_2\sqrt{3})
~~{\rm (lower\ curve)}$$
 is shown together with 
$$ 2 \langle A \rangle / (N_2 \sqrt{3}\, \langle x_l \rangle^2)~~
{\rm (upper\ curve)}$$
for various values of the cosmological constant.
There are almost no finite size effects for these quantities. Within
the accuracy of the plot our data from $8\times 8$, $16\times 16$ and
$32\times 32$ lattices fall on top of one another. The curve for
$\langle \tau \rangle$ decreases for positive values and tends towards 
zero for increasing bare cosmological constant ($\mu \to \infty$),
indicating a crumpled lattice consisting of collapsed triangles.
The ratio is finite for negative values of the cosmological constant
and approaches for $\mu \to \mu_c = -\sqrt{3}$ its maximum
possible value $1$, indicating the dominance of equilateral triangles 
in a locally flat lattice as predicted by equations~(\ref{mu_c}) 
and~(\ref{tau_mu}). Link length fluctuation are demonstrated by the 
fact that the curves for $\langle \tau \rangle $ and 
$2 \langle A \rangle / (N_2\sqrt{3}\langle x_l \rangle^2)$ 
are quite distinct.

\begin{figure}[-t] \begin{center}
\epsfig{figure=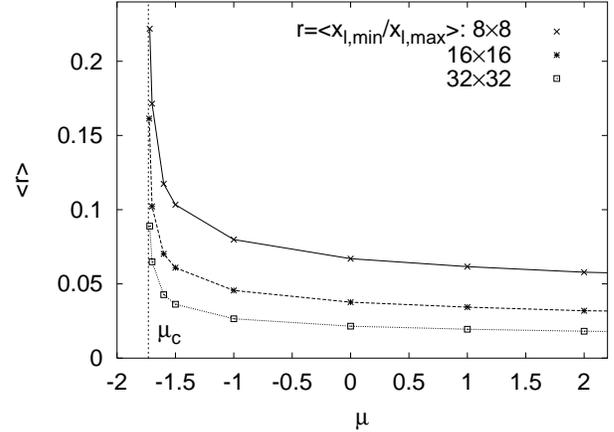,width=\columnwidth} \vspace*{-1mm}
\caption{The expectation value of the ratio $r=\min_l(x_l)/\max_l(x_l)$, 
where minimum and maximum are defined with respect to the configuration 
at hand, is plotted versus $\mu$ for lattice sizes $8\times 8$, 
$16\times 16$ and $32\times 32$.  \label{fig_rat} }
\end{center} \end{figure}

On a single configuration the range of link fluctuations is between
$x_{l,\min}=\min_l(x_l)$ and $x_{l,\max}=\max_l(x_l)$, which define
the minimum and maximum link length on the configuration. In 
figure~\ref{fig_rat} we plot the average value
$$ \langle r \rangle = \langle x_{l,\min}/x_{l,\max} \rangle $$
for our three lattice sizes. As we assumed, the 
expectation value of this ratio decreases with
increasing lattice size and the results
show that even a modest lattice size ($32 \times $32)
is sufficient for substantial variations of the overall
geometry near the critical point ($ \langle r \rangle < 0.1$ for
$\langle \tau \rangle > 0.9$ ).

We are now confident that the assumptions of the previous sections are
essentially correct, leading to the picture of a lattice with highly
correlated links near the critical point.
In the following we check this conclusion,
investigating the expected long-range correlations directly. 
We employ that we know the exact value~(\ref{mu_c})
of $\mu_c$ to control the approach to the critical point. To improve
the numerical stability we add a $n=3$ term $f x_l^3$ and investigate 
the partition function
\begin{eqnarray} \nonumber
Z(\mu,f) = \int_0^\infty \prod_l dx_l \times ~~~~~~~~~~~ \\
 \exp{ \left( - \mu \sum_t A_t[x] -
              \sum_l( x_l^2 + f x_l^3) \right) }\ .
\label{eq13} \end{eqnarray}
We take $\mu=\mu_c$ and are interested to establish 
critical behavior for $f\to 0$. As long as $f>0$, the integral
(\ref{eq13}) is well-defined for all  $\mu$. However, for $f=0$ the
lattice has a finite area for $\mu > \mu_c$ only, as discussed
in subsection~\ref{neq2}.

We compute the
link-link correlation functions defined by
\begin{equation} \label{C_l}
C_l(t) = { \left( \langle x_0 x_t \rangle -
                  \langle x_0 \rangle \langle x_t \rangle \right)\over
           \left( \langle x_0 x_0 \rangle -
                  \langle x_0 \rangle \langle x_0 \rangle \right) }
\end{equation}
where, using the regular triangulation \cite{flat}, two face to face
links pointing in the $\hat{x}$ direction and separated by $t$ lattice
spacings in the $\hat{t}$ direction are correlated. Note that $t$ is
not necessarily proportional to the geometrical distance between the
links, especially for large $t$, since we know that the geometry
fluctuates according to our measurement of $\langle r \rangle$.

\begin{figure}[t] \begin{center}
\epsfig{figure=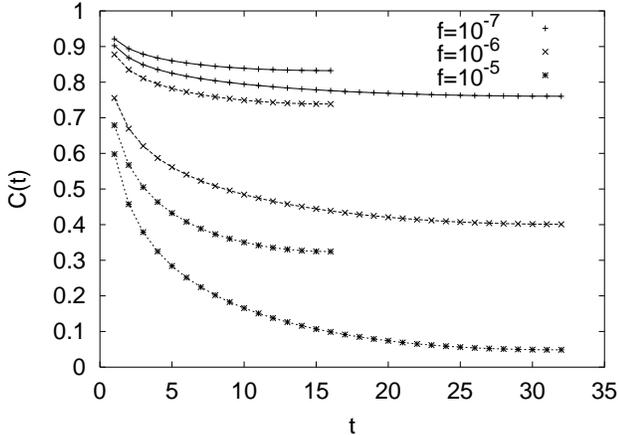,width=\columnwidth} \vspace*{0mm}
\caption{The 2-point functions~(\ref{C_l}) on $32\times 32$ ($t\le 16$)
and $64\times 64$ ($t\le 32$) lattices are plotted for $f=10^{-7}$,
$10^{-6}$ and $10^{-5}$. \label{fig_cor2d} }
\end{center} \end{figure}

Figure~\ref{fig_cor2d}
compares our $32\times 32$ and $64\times 64$ results and
shows indeed a strong correlation of the link lengths for $f \to 0$,
even for large $t$. However, irregular finite size effects are 
present as well (e.g. for $f=10^{-6}$) and a full-scale FSS analysis,
which is beyond the scope of this paper,
seems necessary to clarify the critical behavior of our system.

\begin{figure}[t] \begin{center}
\epsfig{figure=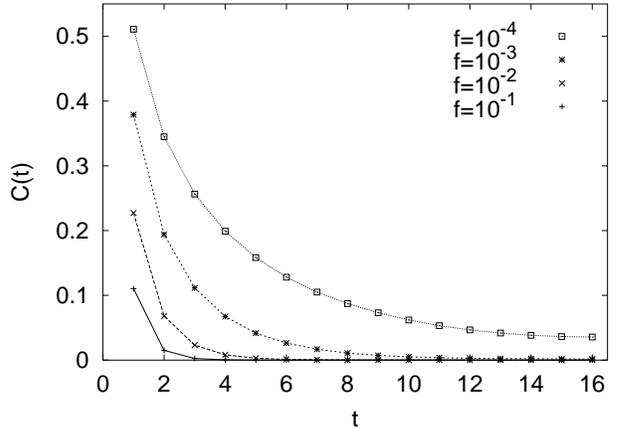,width=\columnwidth} \vspace*{0mm}
\caption{The 2-point functions~(\ref{C_l}) on a $32\times 32$ lattice 
are plotted for our higher $f$ values, $f=10^{-4}$,
$10^{-3}$, $10^{-2}$ and $10^{-1}$. \label{fig_cor2d32b} }
\end{center} \end{figure}

For the $32\times 32$ lattice we performed also calculation at higher
values of $f$. The long distance correlations disappear then quickly,
as is we show in figure~\ref{fig_cor2d32b} (note the change in ordinate
scale between figure~\ref{fig_cor2d} and~\ref{fig_cor2d32b}).
 
\begin{figure}[t] \begin{center}
\epsfig{figure=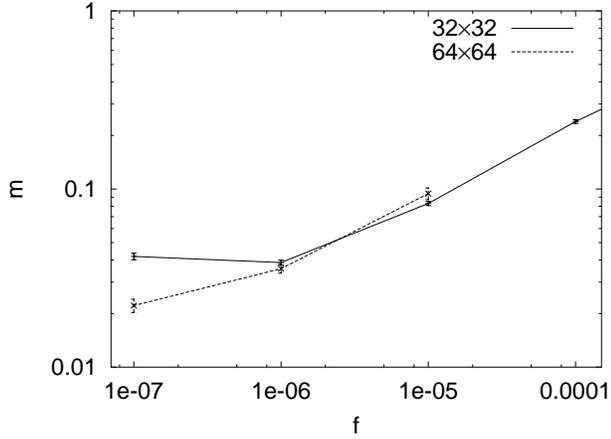,width=\columnwidth} \vspace*{0mm}
\caption{The mass fits (\ref{mass}) from $32\times 32$ and
$64\times 64$ lattices in the limit $f\to 0$. \label{fig_mass} }
\end{center} \end{figure}

In order to determine the mass of the 'particle' associated with
the lattice fluctuations, 
we extracted mass values using so called 
$\cosh$ fits
\begin{equation} \label{mass}
C_l(t) = A\, \left( e^{-mt} + e^{-m(N-t)} \right)
\end{equation}
from the correlation functions. Figure~\ref{fig_mass} depicts our
results from the $32\times 32$ and $64\times 64$ lattices in the limit
$f\to 0$, including also $f=10^{-4}$ from our larger $f$ values on the
$32\times 32$ lattice.  The mass values agree down to 
$f=10^{-6}$; for $f=10^{-7}$ we find then disagreement, which is
expected when the small lattice can now longer accommodate the
correlation length. Again, we emphasize that our data are currently
not sufficient for a complete FSS analysis of the model.
 
\begin{figure}[t] \begin{center}
\epsfig{figure=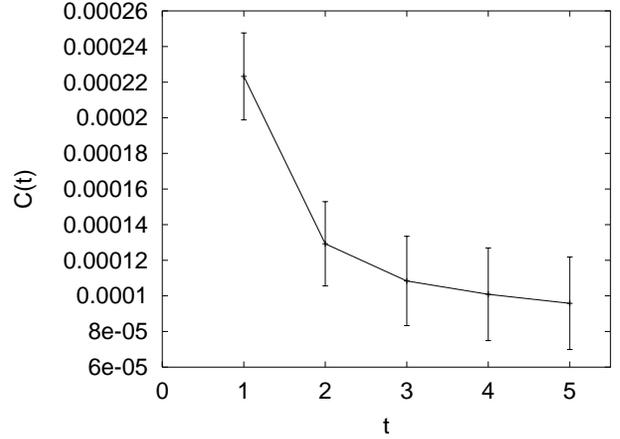,width=\columnwidth} \vspace*{0mm}
\caption{ The 2-point functions~(\ref{C_l}) is shown on a $4^3\,10$
lattice for $\beta=0$, $\mu = -26.5$ and  $\epsilon = 0.01$, where the
correlations are measured along the long direction. \label{fig_cor4d} }
\end{center} \end{figure}

After we have demonstrated numerical results in two dimensions,
we are ready to have a peep at higher dimensions.
Figure~(\ref{fig_cor4d}) shows very
preliminary results of numerical calculations in four dimensions.
We performed these calculations without gravitational coupling (i.e.
$\beta = 0$) near $\mu = \mu_c$ and include a term
$ \epsilon \sum_l q^3_l $
in the partition function (\ref{Zq4d}) for improving convergence. 

The figure shows the results with error bars from two independent
simulation runs on a $4\times 4\times 4\times 10$ lattice, measuring
the correlations for the long direction. Each simulation performed
10k sweeps through the lattice and sampled every 10th configuration.
Depicted is the correlation function 
\begin{equation} \label{C_q}
C_q(t) = { \left( \langle q_0 q_t \rangle -
                  \langle q_0 \rangle \langle q_t \rangle \right)\over
           \left( \langle q_0 q_0 \rangle -
                  \langle q_0 \rangle \langle q_0 \rangle \right) }
\end{equation}
for the links forming the edges of the hypercubes in the lattice. The
correlations are of course measured in the long direction.
Although the figure suffers from large statistical errors, it
exhibits the desired long-range correlations.
 
\section{Conclusions} \label{conclusions}

We defined a class of models of fluctuating geometries, based 
on the Regge approach, to illustrate basic properties of the Polyakov 
path integral in two dimensions. Although we used a simple,
local measure we find evidence for the existence of a non-trivial phase
transition with long-range correlations and the emergence of a scalar 
field.  We demonstrated the importance of a negative bare cosmological 
constant, which may open the door to investigations of related 
systems in higher dimensions. Our calculations indicate that one 
could expect a non-trivial phase transition with long-range 
correlations in four dimensions as well and numerical simulations
with negative cosmological constant in higher dimensions are
the obvious next step to investigate these models further.

\acknowledgments
This work was in part supported by the US Department of Energy under
contract DE-FG02-97ER40608. The simulations were done on workstations
of the FSU HEP group.

\end{document}